# Graph-based Preconditioning Conjugate Gradient Algorithm for "N-1" Contingency Analysis

Yiting Zhao [a,b], Chen Yuan[a], Guangyi Liu[a], Ilya Grinberg[b]
[a] GEIRI North America, San Jose, CA, USA
[b] University at Buffalo, Buffalo, NY, USA
guangyi.liu@geirina.net

*Abstract*—Contingency analysis (CA) plays a critical role to guarantee operation security in the modern power systems. With the high penetration of renewable energy, a real-time and comprehensive "N-1" CA is needed as a power system analysis tool to ensure system security. In this paper, a graph-based preconditioning conjugate gradient (GPCG) approach is proposed for the nodal parallel computing in "N-1" CA. To pursue a higher performance in the practical application, the coefficient matrix of the base case is used as the incomplete LU (ILU) preconditioner for each "N-1" scenario. Additionally, the re-dispatch strategy is employed to handle the islanding issues in CA. Finally, computation performance of the proposed GPCG approach is tested on a real provincial system in China.

## I. Introduction

Contingency analysis is a critical task for power system security to determine the effect of potential contingencies [1]. With the rapid development of smart grid technologies, large amounts of renewable resources have been integrated into the power grid [2], which increases system complexity. It challenges the current data management method and power flow solutions. To guarantee the power system reliability, high-performance computation is much needed for contingency analysis. Most contingency analysis tools utilize the power flow simulation to evaluate the impacts of the outage of a transmission line or transformer. Then the power flow solution is analyzed for overload or branch violations in next step. In the application of contingency analysis, the critical issue is the heavy computation. Reference [3] reduces the number of test scenarios by decision theory. On the other hand, improving the efficiency of power flow computation is another feasible way. In the state of art of the contingency analysis, fast decoupled power flow (FDPF) is a widely used method which adopts LU factorization, followed by forward and backward substitution (F&BS) to solve problem [4] [5]. However, for "N-1" CA scenarios, it has to re-build and re-factorize the coefficient matrices, resulting in a large amount of time consumption. Google Pregel posted the graph idea to "think like a vertex" [6] and storage the data into the graph-based large-scale system to improve performance. In reference [7], it adopts graph theory for network flow algorithm to reduce the calculation time significantly. The graph-based high-performance computing has been introduced to power system analysis [8], which utilizes the constructed relationship among the data by graph database before retrieving request. In this paper, unifying the graph database and iterative method is implemented to achieve the goal of the real-time contingency analysis.

In this paper, a graph-based power system modeling is presented. Based on the graph model, the nodal parallel computing is implemented by the preconditioning conjugate gradient (PCG) algorithm. The graph-based preconditioning conjugate gradient (GPCG) approach is applied to "N-1" contingency analysis (CA) for quick screening. Because of the tiny change from the base case, the coefficient matrix in the base case can be used as incomplete LU (ILU) preconditioning for contingency case to speed up the convergence. For islanding issues in "N-1" scenarios, this paper employs the graph traversal to rapidly detect such situations and to apply the re-dispatch strategy. In this way, it can be efficiently implemented in the graph model to quick screening all "N-1" CA scenarios.

This paper provides a brief introduction to a graph database and the graph modeling of power system in Section II. Section III presents the PCG and elaborates how it be implemented by nodal parallel. In Section IV, it utilizes the base case system states as the ILU preconditioning for "N-1" CA scenarios, and re-dispatch strategy for islanding. Finally, case study and results are given in Section V, and conclusions are summarized in Section VI.

## II. Graph Database and Its Applications for Power Systems

### A. Graph Database and Graph Computing

The graph is a collection of vertices and edges. It is represented as $G = (V, E)$, where $V$ indicates a set of vertices within the graph $G$, and the set of edges is represented as $E$, expressing how these vertices relate to each other. Each edge is

This work was supported by State Grid Corporation technology project SGSHXT00JFJSI700138.

denoted by $e = (i,j) \in E$, where $i \in V$ and $j \in V$ are referred to as the head and tail of the edge $e$, respectively. Graph database is used to model connectivity between objects. Different from the relational database, graph database is the graphic view of the physical data, as shown in Figure 1. Each vertex or edge represents a component in a graph database. The graph structure is consisted of these components and their relationships. For graph database, the graph structure is constructed, during the data loading. Once one entity retrieved, the related components are also detected by one simple operation. Thus, the efficiency of data retrieved, data update and data communication are greatly enhanced by using the graph database.

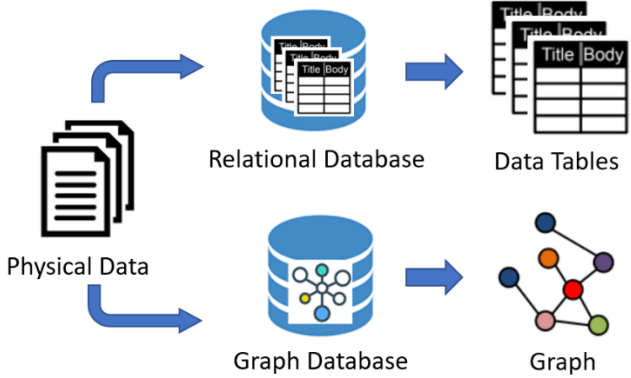

Figure 1. Different view of relational database and graph database

Each vertex in the graph acts as a parallel unit of storage and computation simultaneously [9]. For the storage function, it can save its attributes at local, and directly apply them to the calculation for itself without data communication. Instead of the static data unit, each vertex can be associated with a compute function. To implement nodal parallel calculation, "Local" vertices can send and receive messages to their "Neighbors", which are connected vertices via the related edges in the graph. This local-neighbor pattern will be vividly presented by the example of power balance equation in Section II. C. Once activated, node-based parallel computing is employed by the Bulk Synchronous Parallel (BSP) model. BSP is a bridging model for designing parallel algorithms, which is consisted of components capable of local memory transactions, a network that communicates messages between components and a facility allows for components synchronization [10]. Within the graph processing engine, each processor is assigned to a data partition and equipped with resources to works on its own local computation task. Communication with other processors and outputs during the barrier synchronization are implemented in BSP. In this way, nodal parallel computing is realized by activating nodes at the same time, which ensures the efficiency of graph computing. In this paper, the nodal parallel computing is specifically applied to Section II. C and Section III. B.

*B. Graph-based Power System Model*

Generally, a traditional power system is represented via bus-branch standardized model, which is standardized through IEEE common data format (CDF) [11] [12]. The bus-branch model can be directly mapped to vertex and edge in a graph. Each bus is described as a vertex in the graph model. Similarly, the branch is represented as a bi-directional edge in the graph model of the power system. Then, all parameters of the bus and branch are stored as attributes in the corresponding vertices or edges, including voltage magnitude, voltage angle, power injection, bus type, etc. as vertex-attributes, and transmission line power flow, power flow limits, transformer turns ratio as edge-attributes, as shown in Figure 2. The graph model of the IEEE118-bus system is shown in Figure 3. Furthermore, the graph model is dynamic to update operating status for power flow analysis, like the real-time power injection and the branch connect status for power system analysis.

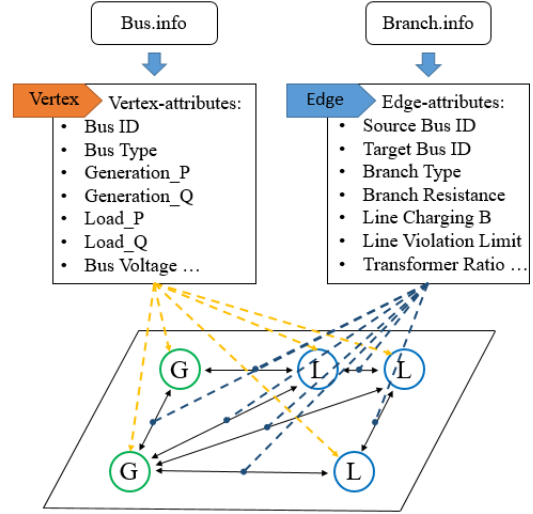

Figure 2. Example of Vertex-Edge Mapped to Bus-branch

*C. Graph-based Nodal Parallel Computing in Power Balance Calculation*

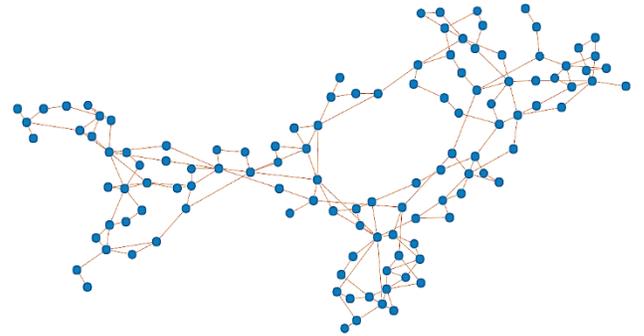

Figure 3. Graph model of IEEE 118-bus system

In this section, graph-based nodal parallel computing is the foundation to implement the power flow calculation in power systems. Bus power injections can be completed with local and neighboring parameters via connected edges.

$$\Delta P_i = P_{is} - |V_i| \sum_{j=1}^{n} |V_j|(G_{ij} \cdot cos\theta_{ij} + B_{ij} \cdot sin\theta_{ij}) \quad (1)$$

Taking the power balance equation as an example, as shown in (1). On the right-hand side, $P_{is}$ is local variable which stands for power injection from bus ($i$). The second summation for each bus can be broken into $P_{i-line}$, the power flow injected to bus ($i$) by connected transmission line, and power cost $|V_i|^2 \cdot G_{ii}$ at bus itself, as shown in equation (2) and (3).

$$\Delta P_i = P_{is} - P_{i-line} - |V_i|^2 \cdot G_{ii} \quad (2)$$

$$P_{i-line} = |V_i| \sum_{j=1, j \neq i}^{n} |V_j|(G_{ij} \cdot cos\theta_{ij} + B_{ij} \cdot sin\theta_{ij}) \quad (3)$$

As attached attributes, bus voltage magnitude and angle are available at vertices $v_i$ and $v_j$, and bus transmission line parameter $G_{ij}, B_{ij}$ has been attached at $e(i,j)$. For the calculated power flow injection from each branch to bus ($i$), the neighbor vertex-attributes $|V_j|, \theta_j$ and edge-attributes $G_{ij}, B_{ij}$, in (3) can be obtained by traversal neighbor vertex $v_j$ and the corresponding edge $e(i,j)$ in graph model. Similarly, for the each local vertex, the $\Delta P_i$ can be independently calculated via native vertex attributes and accumulated transmission line power flow $P_{i-line}$. Taking bus (1) as a local vertex example in the 5-bus system shown in Figure 4, $v_1$ collects the parameters $|V_j|, \theta_j$ from neighbor vertices $v_2, v_3, v_4, v_5$ via edges $e(1,2), e(1,3), e(1,4), e(1,5)$. Along the connected edges, the red arrows represent the request sent by local vertex $v_1$ to its neighbors, and return the voltage magnitude and angle to $v_1$ along black arrows to implement $P_{i-line}$ calculation. Using BSP, the neighbors are activated and processed synchronously. Then, the $\Delta P_i$ can be calculated via local bus attributes, since the node is equipped with storage and computation function simultaneously. In this way, each bus can be independently set as local vertex to retrieve its neighbors to implement the node-based parallel computing in graph based power system model.

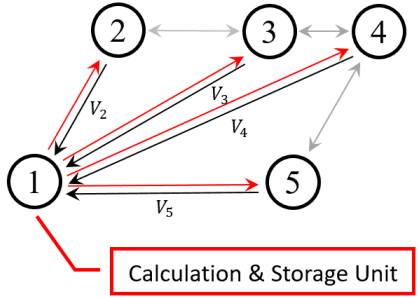

Figure 4. A parallel unit of storage and computation

### III. GRAPH-BASED CONJUGATE GRADIENT AND ITS PRECONDITIONER

#### A. Conjugate Gradient and its Preconditioner

In mathematics, conjugate gradient algorithm is widely employed as an iterative technique to solve sparse symmetric positive definite linear systems [13]. The convergence rate of iterative linear solvers increases as the condition number of the coefficient matrix decreases. Thence, a well-selected preconditioner $M$ is very beneficial in iterative methods to efficiently solve a linear system $A \cdot x = b$. If using preconditioner $M = A^{-1}$, the solution can be obtained directly. However, there is extra cost caused by multiplication by preconditioner matrix $M$ at each step. So, the choice of suitable preconditioner is the trade-off between enhanced convergence and costly multiplication.

**Preconditioning Conjugate Gradient Algorithm**
1. Initial $x_0 = 0$, $r_0 := b - A \cdot x_0$, $z_0 := M^{-1} * r_0$, $p_0 := z_0$,
2. For $k = 1,2,3,\ldots$
3. $\alpha_k = r_{k-1}^T z_{k-1} / p_{k-1}^T A\ p_{k-1}$; Step Length
4. $x_k = x_{k-1} + \alpha_k \cdot p_{k-1}$; Approx. Solution
5. $r_k = r_{k-1} - \alpha_k \cdot A \cdot p_{k-1}$ Residual
6. End if $r_k$ is sufficiently small, then exit loop.
7. k++;
7. $z_k := M^{-1} \cdot r_{k-1}$;
8. $\beta_k = z_k^T r_k / z_{k-1}^T r_{k-1}$; Improvement
9. $p_k = z_k + \beta_k \cdot p_{k-1}$; Search Direction
10. End for-loop;

#### B. Graph-based Preconditioning Conjugate Gradient

In this section, the PCG algorithm is applied in graph computing. According to the local-neighbor pattern for nodal parallel computing, variables in PCG are classified as local variables and neighboring ones in Figure 5. And the step length and improvement are global variables to modify all vertices in the graph. In this way, GPCG is implemented by nodal parallel computing in the graph.

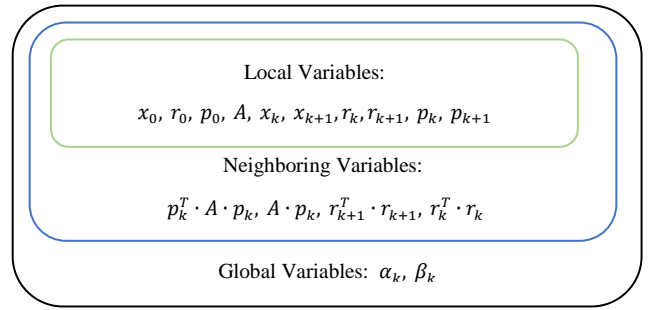

Figure 5. Variable classification of GPCG approach

The local variable is available in the native storage unit, and its update is implemented at its native calculation unit. A neighboring variable needs vertex-attributes by retrieving connected edges and neighbors. Different with local and neighboring variables, the global variables need to collect the summary information, then analysis the improvement and step length for next system update. In this way, GPCG can be re-written as below, where $.BIJ$ is the non-zero element in coefficient matrix $A$, including off-diagonal elements stored as the edge-attributes for edge $e.BIJ$ and diagonal elements attached as vertex-attributes for vertex $s.BIJ$. The local variables are calculated during vertex retrieve and neighboring variables are analyzed by edge traversal. After graph updated, the global variables can be summarized for next iteration.

**Graph-based Preconditioning Conjugate Gradient Algorithm**
1. Initial $x_0 = 0$, $r_0 := b - A \cdot x_0$, $z_0 := M^{-1} * r_0$, $p_0 := z_0$
2. For $k = 1,2,3,\ldots$
3.    *T0: Select an edge e from vertex s in T0 to vertex t*
4.    *Where (shield disconnected edge)*
5.    $s.TempAP += e.BIJ * t.P_{k-1}$,
6.    $PAP += t.P_{k-1} * e.BIJ * t.P_{k-1}$,
7.    *Select a vertex t in T0*
8.    $t.x_k = t.x_{k-1} + \alpha_k * t.P_{k-1}$,
9.    $t.r_k = t.r_{k-1} - \alpha_k * t.TempAP$,
10.    $t.z_k = t.r_k / t.BIJ$,
11.    $r_k^2 += t.r_k * t.z_k$
12. *End ;*
13. End if $r_k$ is sufficiently small, then exit loop;

```
14.     β_k = r_k^2 / r_{k-1}^2;
15.     k++;
16.     T0:  Select a vertex s in T0
17.          s.TempAP = 0,
18.          s.P_k = s.z_k + β_k · s.P_{k-1},
19.          s.TempAP += s.BIJ * t.P_k,
20.          PAP += s.BIJ * s.P_k
21.     End;
22. End for-loop;
```

Therefore, GPCG has high parallelism to solve linear system $A \cdot x = b$, which is applied in power system analysis to solve power flow equation (4) and (5).

$$\Delta P/V = B' \cdot \Delta\theta \quad (4)$$

$$\Delta Q/V = B'' \cdot \Delta V \quad (5)$$

During testing, it is not hard to find that the GPCG convergence highly depends on the initial input vector $x_0$. If $x_0$ is close to the final solution, it will converge quickly. In the next section, GPCG is applied to power systems contingency analysis. Based on the assumption that in "N-1" CA, the new steady state does not deviate far away from the base case solution. So, the base case stats could be employed as the initial input of CA, and the convergence performance is significantly improved.

## IV. "N-1" CONTINGENCY ANALYSIS APPLICATIONS

### A. "N-1" Contingency Analysis in Power Systems

Contingency analysis is necessary to maintain secure system operation such that it can withstand at least one component loss without major interruptions or system violations. It is a "what if" scenario simulator that evaluates, provides and prioritizes the impacts on the power system when outages happen. The contingency is defined as the loss or failure of a single power system component, e.g. a transmission line or a transformer. With a large amount of power components in a given system, "N-1" CA generates the same scale of scenarios for security analysis, as shown in Figure 6. Besides, branch active power flow is major evaluation parameter, so it is focused on equation (4) in quick screening. Instead of the repeating processes of the coefficient matrix for intricate "N-1" CA, GPCG directly calculates the new solution based on base-case system status. Therefore, based on graph computing, PCG can be more efficiently implemented to identify severe branch violation and critical scenarios among "N-1" contingencies.

### B. ILU Preconditioner in "N-1" Scenarios

The reason why PCG selected is that the "N-1" scenarios are similar to the prior base case. It means that $B'_{CA}$ is only slightly deviated from $B'_{basecase}$. According to the "N-1" assumption, where only one component is selected as the outage element, there are only four elements within the matrix changed from prior base case, namely, two diagonal elements $A_{ii}, A_{jj}$, and two off-diagonal elements $A_{ij}, A_{ji}$, as shown in Figure 7. It is proved that $B'_{basecase}$ is highly similar with the $B'_{CA}$ in "N-1" CA scenarios. Therefore, $B'_{basecase}$ is adopted as ILU preconditioning for the $B'_{CA}$ in "N-1" CA scenarios by GPCG.

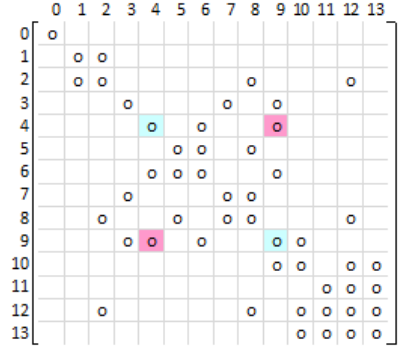

Figure 7. The 4-element difference in IEEE 14-bus system

In the ILU preconditioning phase in each scenario, it solves the equation (6) for $\Delta\theta_{pre}$, where $\Delta P_{CA}$ is calculated in each "N-1" CA scenario, and $|V_{basecase}|$ is solved in base case. Next step, the approximate solution $\Delta\theta_{pre}$ will be corrected by $B'_{CA}$ in equation (7). As the increasing size of power system, it higher similarity between each $B'_{CA}$ and $B'_{basecas}$. In other words, the preconditioning solution $\Delta\theta_{pre}$ is closer to final solution $\Delta\theta_{PCG}$. As result, it will convergent rapidly to correct the solution by $B'_{CA}$.

$$\Delta P_{CA} / |V_{basecase}| = B'_{basecase} \cdot \Delta\theta_{pre} \quad (6)$$

$$\Delta P_{CA} / |V_{solve}| = B'_{CA} \cdot \Delta\theta_{PCG} \quad (7)$$

In the whole process of "N-1" CA, the $B'_{basecase}$ can be stored in graph, and re-used in each scenario graph to reduce cost in parallel processing.

### C. Re-dispatch for Islanding

For islanding scenarios in CA, if the edge is disconnected in graph, it causes single vertex or multiple vertices isolated. During the graph traversal, islanding is detected and analyzed. It reports the power supply exported to or imported from the main power system and the number of generators or loads. For a better convergence performance, the power supply or consumption from separated vertices is compensated by the main power system. The re-dispatch strategy is proportional re-distribution among major generators or loads in the remaining main power system. In the graph-based power system model, the re-dispatched buses can be ranked and filtered by graph traversal. So, the capacity of GPCG is enhanced by the re-dispatch process.

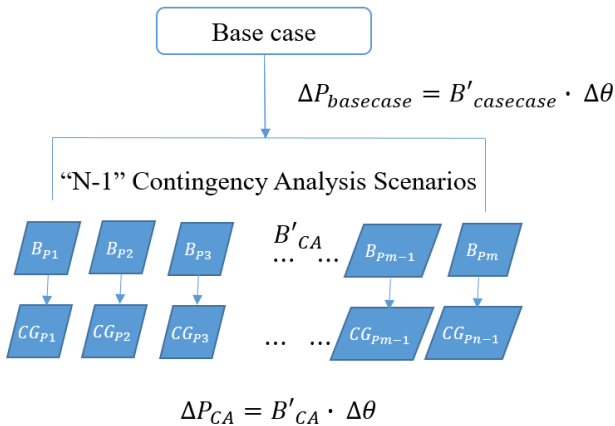

Figure 6. "N-1" CA scenario graphs

## V. CASE STUDY

In this section, GPCG results in "N-1" CA screening are presented. TigerGraph [14] serves as the graph simulation platform. To verify the proposed approach and test its performance, a real provincial system, FJ case, is used. It consists of 1425 buses and 1691 branches. Its graph model is shown in Figure 8. Blue dots represent buses in the power system, and lines, the connected edge between blue dots, present transmission lines.

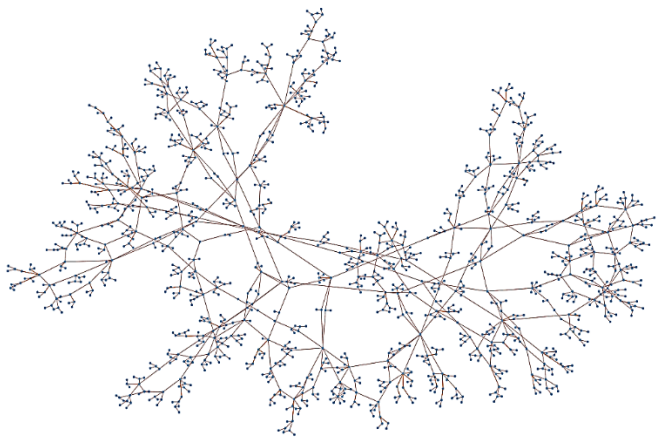

Figure 8. The graph model of FJ case

Totally, 1691 "N-1" contingency scenarios are tested, as displayed in Table I. The LU decomposition is tested as a reference. Because of the repeating processes in each "N-1" CA scenarios, the fast-decoupled method consumes more resources and time. Additionally, the LU decomposition is failure, which needs an extra process for singular matrix caused by the islanding scenarios, while GPCG method can directly solve islanding cases. At present, GPCG can solve all the "N-1" CA scenarios by the simple re-dispatch operation for islanding scenarios.

TABLE I. SCREENING RESULT FOR CONTINGENCY ANALYSIS

| FJ Case | | GPCG | LUD |
|---|---|---|---|
| Tested Branches | | 1691 | 1691 |
| Test Scenarios | Converge | 1691 | 1000 |
| | LU Failure | 0 | 691 |
| | Total | 1691 | 1691 |

Table II lists 5 representatives of "N-1" CA scenarios and the computation time is compared with the fast-decoupled LU decomposition method which abridged as LUD in the table. The testing is conducted on a server with Intel(R) Xeon(R) CPU E7-4830 v3 @ 2.10GHz. The first three test scenarios stand for the 59.14% of "N-1" scenarios. It reports the GPCG is faster than LUD method. The power flow solution time varies depending on the outage component. Caused by the power re-dispatch, the time consumption will increase as the more CG iterations. Last two examples list the LU failure scenarios, which accounts for 40.86% of total CA cases. Totally, the GPCG spent 4.89s on screening all 1691 "N-1" CA scenarios in FJ case.

TABLE II. TIME-CONSUMPTION FOR "N-1" CA SCENARIOS

| | FJ Case | | Time (ms) | |
|---|---|---|---|---|
| # | Branch Outage | Islanding | GPCG | LUD |
| 1 | 5-356 | No | 8.21 | 12.71 |
| 2 | 248-234 | No | 3.18 | 12.10 |
| 3 | 824-131 | No | 10.99 | 13.36 |
| 4 | 1078-1080 | Yes | 8.51 | — |
| 5 | 1158-1159 | Yes | 16.29 | — |

## VI. CONCLUSION

In conclusion, the proposed GPCG approach is feasible and efficient to quick screening all the "N-1" CA scenarios. Based on the graph structure, graph-based power system model is established in graph database for effective data management and analytics. For the PCG algorithm, graph model enlarges the advantage of graph traversal and its parallelism efficiency. The GPCG approach enables nodal parallel computing to synchronously update vertices for high-performance computing. For graph based power system analysis, the proposed GPCG approach plays a positive role in "N-1" contingency analysis.